\documentclass[a4paper,fleqn]{cas-dc}

\usepackage[numbers,sort]{natbib}
\usepackage{tikz}
\usepackage{DejaVuSans}
\usepackage{textcomp}
\hypersetup{hidelinks}

\usepackage{placeins}
\usepackage{ragged2e}
\usetikzlibrary{arrows.meta,calc}
\usepackage[normalem]{ulem}

\ExplSyntaxOn
\cs_set:Npn \__first_footerline:
  {
    \group_begin:
    \rmfamily\small
    \ifnum\theblind>0\relax\else\__short_authors:\fi
    \group_end:
  }
\cs_set:Npn \__cas_head: { }
\cs_set:Npn \__cas_foot:
  {
    \parbox[t]{\textwidth}
      {
        \rule{\textwidth}{.2pt}\\
        \rmfamily\small
        \__first_footerline:
        \hfill Page~\thepage{}~of~\lastpage
      }
  }
\RenewDocumentCommand \printorcid { }
  {
    \seq_if_empty:NF \g_stm_orcid_seq
      {
        \group_begin:
        \tex_let:D \thefootnote \relax
        \footnotetext{\raggedright\textsc{orcid}(s):\c_space_token
          \seq_use:Nn \g_stm_orcid_seq { ;~ }}
        \group_end:
      }
  }
\cs_set_eq:NN \__first_foot: \__cas_foot:
\ExplSyntaxOff
\let\CASoriginalprintemails\printemails
\renewcommand{\printemails}{
  \begingroup
  \renewcommand{\footnotesize}{\fontsize{7}{8}\selectfont}
  \CASoriginalprintemails
  \endgroup
}
\def\tsc#1{\csdef{#1}{\textsc{\lowercase{#1}}\xspace}}
\tsc{WGM}
\tsc{QE}
\tsc{EP}
\tsc{PMS}
\tsc{BEC}
\tsc{DE}

\begin{document}
\setlength{\emergencystretch}{2em}
\let\WriteBookmarks\relax
\def\floatpagepagefraction{1}
\def\textpagefraction{.001}
\shorttitle{}
\shortauthors{Y. Rodriguez Linares et~al.}

\title [mode = title] {Ab Initio Molecular Dynamics-Guided Machine-Learning Potentials for Mapping Nitric Oxide Interactions with Oxygen-Covered Graphite}

%%% affilations %%%

\affiliation[1]{organization={Univ. Lille, CNRS, UMR 8523 - PhLAM - Physique des Lasers Atomes et Molecules},
                city={Lille},
                postcode={F-59000}, 
                country={France}}

\affiliation[2]{
    organization={Univ. Lille, CNRS, INRAE, Centrale Lille, UMR 8207 - UMET - Unité Matériaux et Transformations},
    city={Lille},
    postcode={F-59000},
    country={France}
}

\affiliation[3]{
    organization={Nextmol (Bytelab Solutions SL)},
    city={Barcelona},
    postcode={08014},
    country={Spain}
}

%%%authors%%%

\author[1]{Yaideny {Rodriguez Linares}}
\ead{yaideny.rodriguez-linares@univ-lille.fr}

\author[1,2]{Gilberto A. {Alou Angulo}}
\ead{gilberto.alou@univ-lille.fr}

\author[1,3]{Samuel {Del Fr\'e}}
\ead{samuel.delfre@nextmol.com}

\author[1]{Maurice Monnerville}
\ead{maurice.monnerville@univ-lille.fr}

\author[1]{Alejandro {Rivero Santamar\'ia}}
\cormark[1]
\ead{alejandro.rivero@univ-lille.fr}

\cortext[cor1]{Corresponding author}

\begin{abstract}
Exploratory ab initio molecular dynamics is used to construct machine-learning interatomic potentials for NO interactions with highly oriented pyrolytic graphite at 33\% bridge-site oxygen coverage. Representative configurations sampled across surface temperatures and incidence energies provide reference energies and forces for training and validation. The selected potential enables relaxation of 300 adsorption configurations and construction of spatial and angular interaction maps. These maps reveal increasingly restricted attractive regions during vertical approach and reduced short-range repulsion upon molecular reorientation. The resulting adsorption geometries and interaction landscapes guide further first-principles sampling toward refining potentials for NO oxidation dynamics.
\end{abstract}


\maketitle

\section{Introduction}

Nitrogen oxides (NO$_x$) are major atmospheric pollutants involved in photochemical smog, acid rain, and the formation of secondary reactive nitrogen species \cite{DeVries2021}. Their heterogeneous conversion at carbonaceous interfaces is therefore relevant both to atmospheric chemistry and to the design of carbon-based materials for pollutant mitigation \cite{HouCen2015}. Graphitic surfaces are commonly used as molecular-scale models of soot-like carbon particles, whose reactivity is strongly affected by defects and oxygen-containing functional groups \cite{Xi2021,HouCen2015}.

Previous density-functional theory (DFT) studies have predicted that NO can be oxidized by epoxy groups on graphene oxide through a reaction pathway with an activation barrier of approximately \mbox{$0.1\,\text{eV}$}, while neighboring hydroxyl groups can modify adsorption and reaction energetics \cite{HouCen2015,Cen2015}. However, identifying a low-energy reaction pathway does not establish whether or how the reaction occurs under dynamical conditions, where surface thermal motion, energy redistribution, and collision geometry can determine whether NO scatters, becomes trapped, or reacts.

To address these dynamical aspects, previous work by some of the present authors employed \textit{ab initio} molecular dynamics (AIMD) to investigate NO collisions with highly oriented pyrolytic graphite (HOPG) functionalized by an isolated epoxy oxygen atom (O@HOPG), with the surface thermalized at \mbox{$300\,\text{K}$}. By considering different incidence energies and molecular orientations, these simulations showed that NO$_2$ formation can occur even at the lowest incidence energy investigated (\mbox{$0.025\,\text{eV}$}), comparable to the thermal energy scale at room temperature \cite{Alou2024}.

A natural question is whether greater surface oxygen coverage enhances NO oxidation by providing more reactive sites. Neighboring oxygen atoms may, however, modify the interaction landscape, accessibility of epoxy groups and short-range repulsion, changing the balance between scattering, trapping and reaction. Whether a highly oxygenated graphite surface promotes or inhibits NO oxidation under collision conditions therefore remains open.

To address this question, the present work adopts a different computational strategy from that previous AIMD study~\cite{Alou2024}, combining exploratory AIMD simulations with machine-learning interatomic potentials (MLIPs). Rather than seeking statistically converged collision outcomes from computationally expensive AIMD trajectories, a limited set of simulations at different surface temperatures and incidence energies is used to sample a diverse range of atomic configurations in the NO--surface interaction space. The resulting first-principles data provide the basis for developing MLIPs capable of describing the interaction between NO and the highly oxygen-covered O$_{16}$@HOPG surface.

These MLIPs are developed using the workflow established by Del Fr\'e et al.~\cite{DelFre2026}, combining representative configuration selection, model training, and validation against DFT calculations. The resulting potentials are used to investigate NO adsorption and construct spatial and angular interaction maps, providing insight into the interaction landscape and accessible molecular configurations. This approach represents a first step toward large-scale molecular dynamics simulations aimed at obtaining statistically meaningful insights into NO oxidation dynamics on oxygen-covered graphite, including the competition between scattering, trapping, and reactive processes, at a significantly reduced computational cost.

\section{Computational Methods}

\subsection{Surface model and electronic-structure calculations}

All electronic-structure calculations were performed with the Vienna Ab initio Simulation Package (VASP) \cite{Kresse1996CMS,Kresse1996PRB} using the projector augmented-wave method \cite{Blochl1994,KresseJoubert1999}. The exchange-correlation energy was described within the generalized gradient approximation using the Perdew-Burke-Ernzerhof (PBE) functional \cite{Perdew1996}, including the DFT-D3 dispersion correction with Becke-Johnson damping \cite{Grimme2010,Grimme2011}. A plane-wave kinetic energy cutoff of \mbox{$500\,\text{eV}$} and a Gaussian smearing width of \mbox{$0.05\,\text{eV}$} were used. The Brillouin zone was sampled with a 5 $\times$ 5 $\times$ 1 Monkhorst-Pack $k$-point mesh \cite{Monkhorst1976}.

The HOPG substrate was modeled by a p(4 $\times$ 4) graphite slab containing three graphene layers in AB stacking, with 32 carbon atoms per layer and 96 carbon atoms in total. Three-dimensional periodic boundary conditions were applied, with a vacuum region of approximately \mbox{$15\,\text{\AA}$} between the uppermost carbon layer and its periodic image along the surface normal. After oxygen functionalization, the effective vacuum distance measured from the top of the oxygen-covered surface was approximately \mbox{$13.8\,\text{\AA}$}.

\begin{figure*}[pos=t!]
\centering
\includegraphics[width=0.80\textwidth]{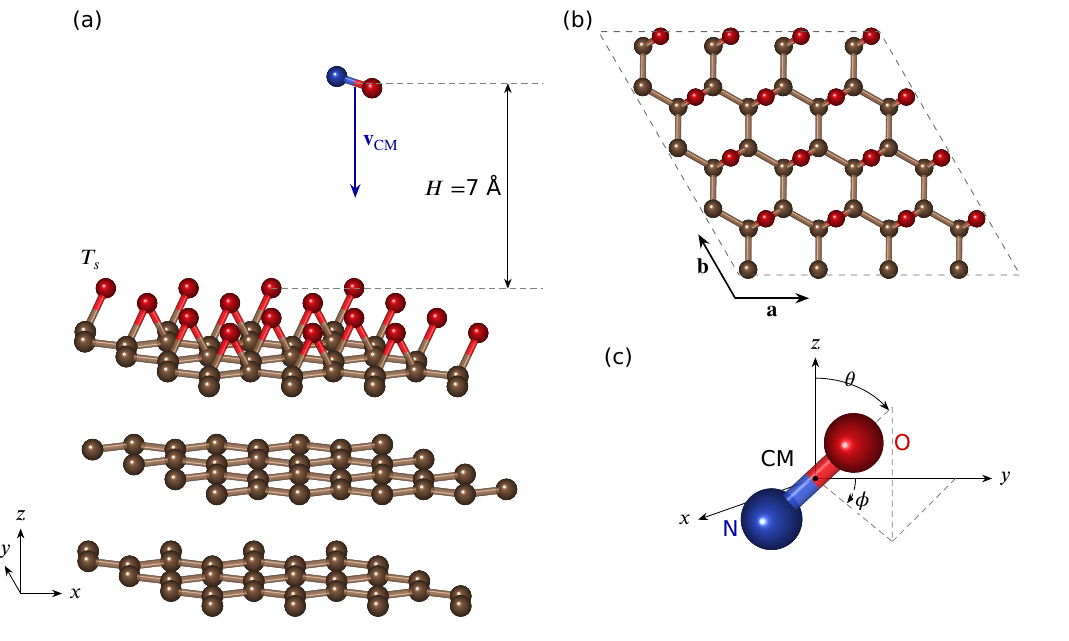}
\par\medskip
\refstepcounter{figure}\label{fig:initial-conditions}
{\small\rmfamily\justifying\noindent\textbf{Figure~\thefigure:} Coordinate system and sketch of the initial conditions used in the AIMD simulations for NO collisions with O$_{16}$@HOPG: (a) incident molecule impinging the surface in normal orientation, initial height $H$, center-of-mass velocity $\mathrm{V}_{\mathrm{CM}}$ and $T_S$ representing the thermalized surface; (b) top view of the surface cell; and (c) molecular orientation defined by $\theta$ and $\phi$ that were randomly generated.\par}
\end{figure*}

The O$_{16}$@HOPG surface was constructed by placing 16 oxygen atoms on bridge sites of the exposed graphite layer (Fig.~\ref{fig:initial-conditions}b). The 32 surface carbon atoms define 48 C--C bridge sites, of which 16 are occupied, corresponding to a 33\% bridge-site coverage. The bridge position was selected because atomic oxygen adsorbed at this site forms a stable epoxide-like configuration on graphite, consistent with previous calculations on oxygen-functionalized HOPG and graphene-based models \cite{HouCen2015,Alou2024}. The oxygen atoms were distributed over the exposed layer so as to form a high-coverage oxidized surface while enforcing a minimum O--O distance of \mbox{$2.0\,\text{\AA}$}.

The slab was relaxed to residual forces below \mbox{$0.01\,\text{eV}\,\text{\AA}^{-1}$} with an electronic convergence threshold of \mbox{$10^{-6}\,\text{eV}$}. The bottom graphite layer was fixed to represent the underlying bulk; all other atoms were relaxed. The optimized slab was the reference for adsorption calculations and AIMD.

\subsection{NO gas-phase reference and NO/NO$_2$ adsorption calculations}
\label{subsec:Adsorption_calc}

Before carrying out the exploratory AIMD simulations, DFT calculations were performed to characterize the isolated NO molecule and its interaction with the optimized O$_{16}$@HOPG surface. The intramolecular N--O interaction was first examined through a spin-polarized point-by-point potential energy scan. In these calculations, the NO molecule was placed far from the oxidized surface, and the N--O distance was varied from 0.8 to \mbox{$4.0\,\text{\AA}$}. This scan was used to validate the description of the NO bond at the same level of theory employed for the surface calculations and to provide the intramolecular potential used to generate the initial conditions for AIMD.

NO adsorption on O$_{16}$@HOPG was then investigated in order to characterize the molecule--surface interaction before the dynamical simulations. Three surface oxygen atoms were selected: the one closest to the cell center and two others nearest to target distances of 1 and \mbox{$2\,\text{\AA}$} from the center. Five initial molecular configurations were considered for each selected oxygen atom, giving a total of 15 adsorption configurations. In the three horizontal configurations, the N--O bond was parallel to the surface, with either the NO center of mass, the nitrogen atom, or the oxygen atom positioned directly above the selected surface oxygen. In the two vertical configurations, the N--O bond was perpendicular to the surface, with either nitrogen or oxygen pointing downward toward the selected surface oxygen. The initial NO configurations were placed \mbox{$2.5\,\text{\AA}$} from the surface oxygen layer. Each configuration was relaxed with and without spin polarization, using matching spin treatments for the adsorbed and far-reference energies. The adsorption energy was evaluated as:

\begin{equation}
E_{\mathrm{ads}} =
E_{\mathrm{NO/O_{16}@HOPG}}^{\mathrm{near}} -
E_{\mathrm{NO/O_{16}@HOPG}}^{\mathrm{far}},
\end{equation}

where $E_{\mathrm{NO/O_{16}@HOPG}}^{\mathrm{near}}$ is the total energy of the relaxed system with NO close to the oxidized surface. $E_{\mathrm{NO/O_{16}@HOPG}}^{\mathrm{far}}$ is the corresponding energy of the relaxed system with NO placed far from the surface. The adsorption results were compared with previous studies of NO adsorption on oxidized graphitic surfaces \cite{Cen2015,Alou2024}.

NO$_2$ adsorption was studied on O$_{15}$@HOPG, obtained by removing the surface oxygen involved in product formation. NO$_2$ was first relaxed \mbox{$7.5\,\text{\AA}$} above the top carbon layer, then placed \mbox{$2.5\,\text{\AA}$} above that layer, directly above the vacancy left by the removed oxygen atom, in three orientations: horizontal N-down, horizontal O-down and vertical. Each configuration was fully relaxed with and without spin polarization. The NO$_2$ adsorption energy was evaluated using the same near--far reference convention as for NO,

\begin{equation}
E_{\mathrm{ads}}^{\mathrm{NO_2}} =
E_{\mathrm{NO_2/O_{15}@HOPG}}^{\mathrm{near}} -
E_{\mathrm{NO_2/O_{15}@HOPG}}^{\mathrm{far}}.
\end{equation}

The near and far references contain, respectively, adsorbed and separated NO$_2$ on the same O$_{15}$@HOPG substrate, evaluated with matching spin treatments. They provide a reference for product retention and desorption.

\subsection{AIMD sampling}
\label{sec:dynamics-methods}

Exploratory AIMD simulations were performed with VASP to sample a diverse range of atomic configurations of the NO/O$_{16}$@HOPG system, providing DFT reference data for the subsequent development of machine-learning interatomic potentials (MLIPs). These simulations were performed without spin polarization to reduce the computational cost, following the previous O@HOPG study \cite{Alou2024}, in which spin-polarized and non-spin-polarized AIMD trajectories showed similar dynamical behavior. Prior to the collision simulations, the oxidized HOPG slab was thermalized at surface temperatures of 100, 200, 300, 400, and \mbox{$500\,\text{K}$} using canonical (NVT) AIMD simulations with the Nos\'e--Hoover thermostat implemented in VASP. Each thermalization trajectory was propagated for \mbox{$10\,\text{ps}$} with a time step of \mbox{$1\,\text{fs}$}. Surface snapshots extracted from the equilibrated thermalization trajectories were then used as initial substrate configurations, thereby accounting for the instantaneous thermal fluctuations of the surface, as usual in gas-surface simulations \cite{DelFre2026,Alou2024,rivero_santamaria_ab_2019}.

The incoming NO molecule was initialized with an internal energy corresponding to its rovibrational ground state (\mbox{$v=0$}, \mbox{$j=0$}) within a quasiclassical framework \cite{TruhlarMuckerman1979}, with the initial N--O distance and internal momentum determined from the DFT potential energy curve described above. For each trajectory, the NO center of mass was initially placed \mbox{$7\,\text{\AA}$} above the highest atom of the surface configuration. The initial conditions are illustrated in Fig.~\ref{fig:initial-conditions}. The initial $x,y$ coordinates of the NO center of mass were randomly sampled within the simulation cell, while the initial translational velocity was directed along the surface normal toward the substrate, corresponding to normal incidence. Incidence energies of 0.1, 0.3, 0.5, 0.7, 0.8, and \mbox{$2.0\,\text{eV}$} were considered. After initialization, the AIMD trajectories were propagated under constant-total-energy conditions (NVE), using a time step of \mbox{$1\,\text{fs}$}. Trajectories were stopped when the molecular center of mass reached a height greater than \mbox{$10\,\text{\AA}$} above the surface or after \mbox{$4\,\text{ps}$}.

\subsection{MLIP training and validation}
\label{sec:mlip-methods}

The atomic configurations generated by AIMD simulations provided the initial dataset for MLIP training. The MLIP was developed following the workflow of Del Fr\'e et al. \cite{DelFre2026}, which combines representative configuration selection, Deep Potential training using \textsc{DeePMD-kit} \cite{Zhang2018DeepPotential,Wang2018DeePMDkit}, adopting the same training setup as in previous studies \cite{InfusoDeepPotential2025,DelFre2026}, and a query-by-committee (QBC) step \cite{Seung1992QBC,Smith2018LessIsMore,Zhang2019DPGEN} to identify additional configurations for first-principles labelling and subsequent enrichment of the training dataset.

The first stage of this workflow addresses the redundancy of molecular dynamics data: consecutive frames of a trajectory are strongly correlated, so that uniform sampling in time produces a dataset far larger than the diversity it actually contains. Each local atomic environment was therefore encoded as a smooth overlap of atomic positions (SOAP) descriptor~\cite{Bartok2013SOAP}, and the resulting feature space was compressed by principal component analysis (PCA)~\cite{jolliffe2016PCA}, retaining the components that account for 95\% of its variance. Farthest point sampling (FPS)~\cite{imbalzano2018FPS} was then applied in this reduced space, and the selected configurations were partitioned into training and validation sets in an 80:20 ratio.

The selected configurations were used to train four Deep Potential models differing only in the random initialization of their weights. Their performance was evaluated using the root mean square errors (RMSEs) of the predicted energies and forces relative to the DFT reference data. The model exhibiting the lowest errors was then selected to generate MLIP-driven molecular dynamics trajectories for QBC sampling using the Large-scale Atomic/Molecular Massively Parallel Simulator (\textsc{LAMMPS}) \cite{Thompson2022LAMMPS} with the DeePMD-kit interface, following the same protocol as the AIMD simulations except for a time step of \mbox{$0.1\,\text{fs}$}. For this purpose, 20 initial conditions were selected from the exploratory AIMD simulations described in Section~\ref{sec:dynamics-methods}, providing physically relevant starting points for the QBC sampling. 

Configurations extracted from these trajectories were then evaluated by the full committee of four models, using the maximum, over all atoms, of the standard deviation across the four models of the predicted force magnitude as a disagreement criterion. Configurations with disagreement scores between 0.05 and \mbox{$0.50\,\text{eV}\,\text{\AA}^{-1}$} were selected for additional single-point DFT calculations with VASP. This selection prioritizes configurations exhibiting appreciable model disagreement while excluding those with extreme disagreements, which may correspond to regions far from the initial training distribution. Further MLIP refinement was considered unnecessary if at least 95\% of the QBC-evaluated configurations exhibited force prediction disagreements below \mbox{$0.05\,\text{eV}\,\text{\AA}^{-1}$}.

\subsection{MLIP applications}
\label{sec:mlip-applications-methods}

The selected MLIP was applied to explore different aspects of the NO--O$_{16}$@HOPG interaction landscape. First, geometry optimizations from diverse initial configurations were used to identify favorable NO adsorption structures. Spatial interaction energy maps were then constructed to examine how the interaction varies across the oxygenated surface as NO approaches it in a fixed orientation. Finally, angular interaction energy maps were used to investigate the influence of molecular orientation on the accessibility of different regions of the configurational space. Together, these complementary approaches provide a more detailed picture of the NO--surface interaction landscape and help identify configurations of interest for further first-principles sampling.

For adsorption sampling, initial configurations were generated with random NO positions, heights, and orientations and subsequently relaxed with the selected MLIP using the Broyden--Fletcher--Goldfarb--Shanno (BFGS) optimizer in the Atomic Simulation Environment (ASE) \cite{Larsen2017ASE}, with only the bottom carbon layer held fixed to represent the underlying bulk, as in the VASP adsorption calculations. MLIP adsorption energies were evaluated using the same near--far convention as the non-spin-polarized DFT calculations, with the far-reference geometry evaluated using the MLIP.

Spatial interaction energy maps were constructed with NO held vertically, with nitrogen pointing toward the surface, to probe the approach of the N atom to surface epoxy oxygen atoms in geometries relevant to NO oxidation. The height of the NO center of mass above the mean height of the surface oxygen atoms was varied from 1.5 to \mbox{$4.0\,\text{\AA}$}. The molecule was translated over an $x,y$ grid above the unthermalized O$_{16}$@HOPG structure, keeping the substrate fixed. At each point, the interaction energy was calculated with the selected MLIP relative to NO at a molecule--surface distance of \mbox{$7.0\,\text{\AA}$}. The surface reference was the mean height of the surface oxygen atoms.

Angular interaction maps were then calculated at the surface positions with the lowest interaction energies in the previously calculated spatial maps. At each selected position, the NO center of mass was held at 1.5, 2.0, or \mbox{$3.0\,\text{\AA}$} above the mean height of the surface oxygen atoms, while the molecular orientation was varied through the polar angle $\theta$, measured from the surface normal (\mbox{$\theta=0^\circ$}: N down), and the azimuthal angle $\phi$. Interaction energies were calculated with the selected MLIP relative to the same molecular orientation with its center of mass at \mbox{$7.0\,\text{\AA}$}.

The 20 LAMMPS trajectories used for the QBC analysis described in Section~\ref{sec:mlip-methods} were also used to assess computational cost.

\section{Results and discussion}

\subsection{Surface structure and static validation of NO and NO$_2$ adsorption}

The optimized O$_{16}$@HOPG slab retains the layered graphite framework, with oxygen atoms bound to bridge sites on the exposed carbon layer. The relaxed C--O bond length is approximately \mbox{$1.44\,\text{\AA}$}, consistent with the range of \mbox{$1.44 \,\text{--}1.48\,\text{\AA}$} reported in DFT calculations for epoxide-like oxygen adsorption on graphite at different coverages \cite{Moron2011}. The average in-plane C--C bond length is approximately \mbox{$1.42\,\text{\AA}$}, in agreement with the experimental value of \mbox{$1.42\,\text{\AA}$} for pristine graphite \cite{Trucano1975}. The average interlayer separation is \mbox{$3.35\,\text{\AA}$}, close to the experimental value of \mbox{$3.36\,\text{\AA}$} for pristine graphite \cite{Trucano1975}. 

The isolated NO molecule was assessed at the same DFT level used for the gas--surface calculations. A Morse fit to the spin-polarized potential energy curve yields an equilibrium bond length of \mbox{$1.15\,\text{\AA}$} and a harmonic vibrational frequency of \mbox{$1949.91\,\text{cm}^{-1}$}, compared with the experimental values of \mbox{$1.15\,\text{\AA}$} and \mbox{$1904.20\,\text{cm}^{-1}$}, respectively \cite{HuberHerzberg1979}. The calculated bond length is also close to the PBE-D3(BJ) value of \mbox{$1.17\,\text{\AA}$} reported previously \cite{Alou2024}. These results support a reasonable near-equilibrium description of the NO bond for generating the semiclassical initial conditions.

\begin{figure}[pos=tbp]
\centering
\includegraphics[width=\linewidth]{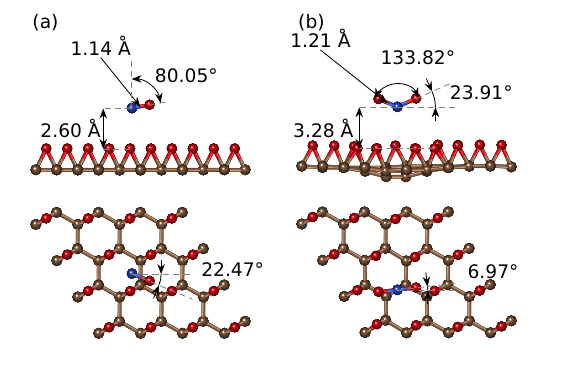}
\par\medskip\refstepcounter{figure}\label{fig:adsorption-geometries}
{\small\rmfamily\justifying\noindent\textbf{Figure~\thefigure:} Optimized DFT adsorption geometries of (a) NO on O$_{16}$@HOPG and (b) NO$_2$ on O$_{15}$@HOPG, in front (upper row) and top (lower row) views. Arrows indicate interatomic distances and molecular angles.\par}
\end{figure}

\begingroup
\emergencystretch=1em
Subsequently, static NO adsorption calculations on O$_{16}$@HOPG were performed, and after relaxation, the horizontal configurations were found to be energetically more stable than the vertical ones. Adsorption energies range from $-0.37$ to \mbox{$-0.20\,\text{eV}$} with spin polarization and from $-0.49$ to \mbox{$-0.30\,\text{eV}$} without it, with modest variation among the reference sites. The spin-polarized minimum, approximately \mbox{$-0.37\,\text{eV}$}, indicates stronger binding than the \mbox{$-0.17\,\text{eV}$} reported for the single-oxygen O@HOPG reference with PBE-D3 \cite{Alou2024} and the \mbox{$-0.10\,\text{eV}$} obtained with PBE-D2 for an epoxy-functionalized graphene oxide model \cite{Cen2015}. The optimized NO bond length is \mbox{$1.14\,\text{\AA}$}, while the N--O$_{\mathrm{surf}}$ separation is \mbox{$2.60\,\text{\AA}$} (Fig.~\ref{fig:adsorption-geometries}a).
\par
\endgroup

Finally, NO$_2$ was characterized on O$_{15}$@HOPG, the substrate remaining after removal of the oxygen atom involved in its formation. Relaxation far from the surface gives an N--O bond length of \mbox{$1.21\,\text{\AA}$} and an O--N--O angle of 133.51$^\circ$, close to the experimental values of \mbox{$1.19\,\text{\AA}$} and 134$^\circ$ \cite{Herzberg1966}. For the three initial adsorption orientations, energies range from $-0.16$ to \mbox{$-0.15\,\text{eV}$} without spin polarization and from $-0.13$ to \mbox{$-0.12\,\text{eV}$} with it. The spread is only \mbox{$0.01\,\text{eV}$} within either treatment; spin polarization weakens binding by \mbox{$0.02\text{--}0.04\,\text{eV}$}. The most stable spin-polarized configuration, obtained from horizontal N-down NO$_2$, retains an N--O bond length of \mbox{$1.21\,\text{\AA}$} and an O--N--O angle of 133.82$^\circ$, indicating little distortion relative to the far-reference molecule (Fig.~\ref{fig:adsorption-geometries}b). Binding is weaker than the \mbox{$-0.26\,\text{eV}$} reported for the single-oxygen reference system \cite{Alou2024}, indicating weaker retention of the product on the O$_{15}$@HOPG surface. Surface deformation is also observed after removal of an oxygen atom. The resulting changes in local bonding induce a reorganization of the neighboring carbon atoms toward a new equilibrium geometry. Specifically, the two carbon atoms formerly bridged by the removed oxygen move downward by approximately \mbox{$0.42\,\text{\AA}$}, while their C--C distance decreases from approximately 1.41 to \mbox{$1.34\,\text{\AA}$}. Such structural rearrangements may influence the subsequent gas--surface dynamics, particularly the interaction of NO$_2$ with the surface following its formation. 

Overall, these calculations provide energetic and structural references for interpreting the four relevant collision outcomes: NO scattering or adsorption on O$_{16}$@HOPG, and NO$_2$ scattering or adsorption on O$_{15}$@HOPG following oxidation.

\subsection{AIMD sampling and collision outcomes}
\label{sec:aimd-results}

The exploratory AIMD dataset comprises 142 collision trajectories over incidence energies of \mbox{$0.1\text{--}2.0\,\text{eV}$} and surface temperatures of \mbox{$100\text{--}500\,\text{K}$}. Rather than aiming at statistically converged reaction probabilities, these simulations were designed to sample diverse atomic configurations under different collision conditions, providing first-principles reference data for subsequent MLIP development. The dominant outcome is non-reactive scattering following interaction with the oxygenated layer. Only one trajectory produces NO$_2$, giving an observed reactive fraction of 1/142 (0.70\%) across the full dataset. This event occurs at \mbox{$E_i=0.3\,\text{eV}$} and \mbox{$T_s=300\,\text{K}$}, where 50 trajectories were sampled and the observed reactive fraction is 2.0\%. The overall fraction combines different conditions and unequal sample sizes; it therefore describes the analyzed dataset rather than a reaction probability at a specific collision condition.

To gain further insight into the collision geometries sampled by AIMD, particularly those potentially relevant to NO oxidation, Fig.~\ref{fig:aimd-trajectory-comparison} compares the minimum distance between the nitrogen atom and a surface oxygen atom reached during each trajectory with the corresponding O--C distance at that instant. Several non-reactive encounters reach short N--O$_{\mathrm{surf}}$ distances while the oxygen remains bound to the substrate, showing that close approach alone is insufficient to trigger reaction. Of the 142 trajectories, 128 reach \mbox{$\mathrm{N}\text{--}\mathrm{O}_{\mathrm{surf}}<2.2\,\text{\AA}$} and 37 reach distances below \mbox{$2.0\,\text{\AA}$}, yet only one produces NO$_2$. Four trajectories reach distances below \mbox{$1.8\,\text{\AA}$} and two below \mbox{$1.6\,\text{\AA}$}. The latter pair comprises the reactive trajectory and a non-reactive encounter at \mbox{$E_i=2.0\,\text{eV}$} and \mbox{$T_s=300\,\text{K}$}. In the non-reactive case, the N--O$_{\mathrm{surf}}$ distance reaches \mbox{$1.41\,\text{\AA}$}, while the corresponding O--C distance at that instant remains \mbox{$1.61\,\text{\AA}$} and the largest O--C distance throughout the trajectory remains below \mbox{$1.77\,\text{\AA}$}. Thus, even a very short contact can occur without oxygen abstraction.

\par\medskip
\noindent\begin{minipage}{\columnwidth}
\centering
\includegraphics[width=0.95\linewidth]{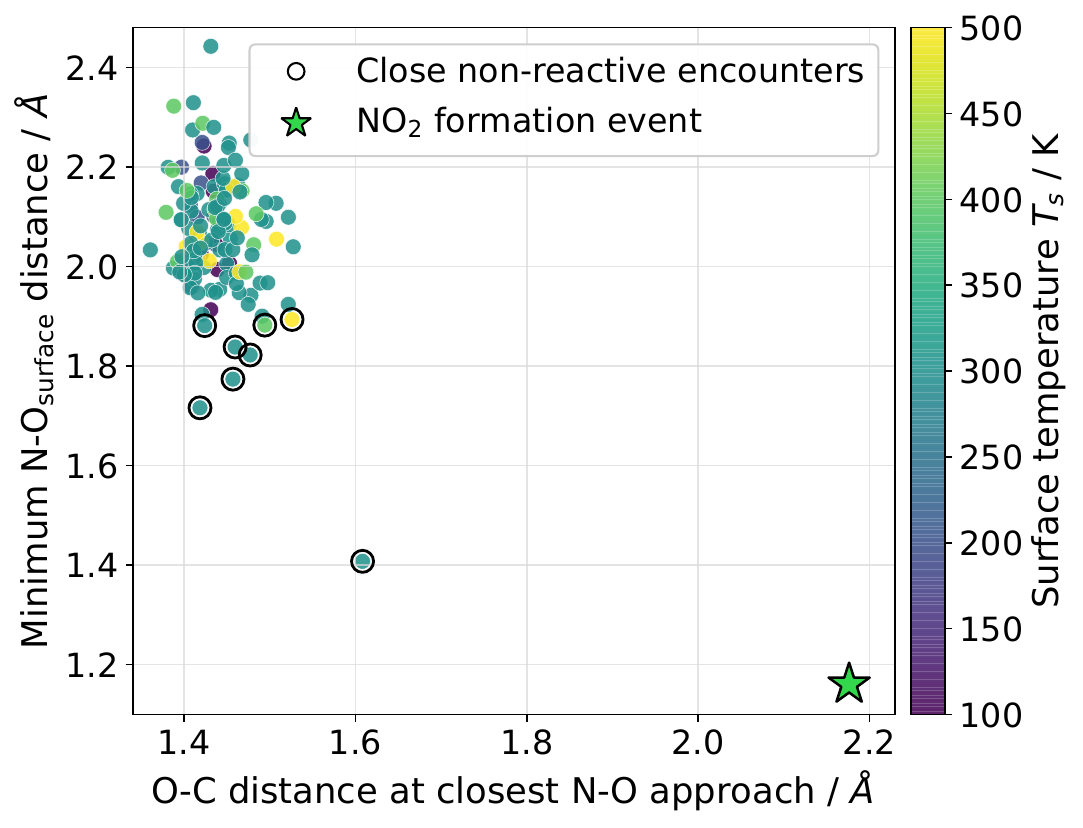}
\par\medskip\refstepcounter{figure}\label{fig:aimd-trajectory-comparison}
{\small\rmfamily\justifying\noindent\textbf{Figure~\thefigure:} Minimum N--O$_{\mathrm{surf}}$ distance and simultaneous O--C distance for each AIMD trajectory. Colours indicate surface temperature. The green star marks NO$_2$ formation; black circles mark non-reactive encounters below \mbox{$1.9\,\text{\AA}$}.\par}
\end{minipage}
\par\medskip

The reactive trajectory follows an indirect sequence: NO first rebounds from the surface and forms NO$_2$ during a subsequent approach. At the closest contact, the N--O$_{\mathrm{surf}}$ distance is \mbox{$1.16\,\text{\AA}$}, close to the equilibrium N--O bond length in NO$_2$, while the O--C distance associated with that oxygen is \mbox{$2.18\,\text{\AA}$}, consistent with oxygen release as the product forms and leaves the substrate. Reaction after the rebound suggests that, in addition to a short N--O distance, a favorable molecular orientation relative to the surface oxygen is needed to promote bond formation. The single reactive event supports this interpretation, although it does not yet establish a unique reactive orientation. 

Neither higher incidence energy nor higher surface temperature produces a systematic increase in NO$_2$ formation within the present sampling. All ten trajectories at \mbox{$2.0\,\text{eV}$} and \mbox{$300\,\text{K}$} remain non-reactive, despite reaching close contacts, and the temperature series at \mbox{$0.3\,\text{eV}$} yields only the event at \mbox{$300\,\text{K}$}. For the single-oxygen O@HOPG model, a reaction probability of 9.0\% was reported at \mbox{$0.3\,\text{eV}$} and \mbox{$300\,\text{K}$} under normal incidence \cite{Alou2024}, compared with the 2.0\% observed here at the same conditions. Although one reactive event is insufficient for a statistically converged comparison, the present results suggest that increasing oxygen coverage may restrict access to configurations that permit oxygen abstraction.

While these exploratory simulations provide insight into the collision dynamics, their computational cost remains substantial, requiring approximately six days per trajectory using 24 CPUs. This reinforces the strategy of using a limited number of first-principles trajectories to sample diverse atomic configurations and develop MLIPs for a more extensive and computationally efficient exploration of the configurational space relevant to NO oxidation on O$_{16}$@HOPG.

\subsection{MLIP training and validation}
\label{sec:mlip-results}

The selected dataset obtained following the protocol developed by Del Fr\'e et al. \cite{DelFre2026}, described in Section~\ref{sec:mlip-methods}, is represented in Fig.~\ref{fig:mlip-fps}. Starting from the 313 816 configurations generated by the exploratory AIMD simulations, farthest point sampling (FPS) in the SOAP/PCA space enabled selection of a representative subset of 2981 configurations (approximately 0.95\% of the original dataset). The figure shows that these span the main branches and less populated regions of the projection, reducing frame redundancy while representing sampled environments. It is worth noting that this selection reflects the configurational space explored by AIMD, where only one reactive event was observed. Consequently, although the selected dataset captures a broad diversity of sampled environments, the extent to which it represents configurations relevant to NO oxidation remains to be established.

\par\medskip
\noindent\begin{minipage}{\columnwidth}
\centering
\includegraphics[width=\columnwidth]{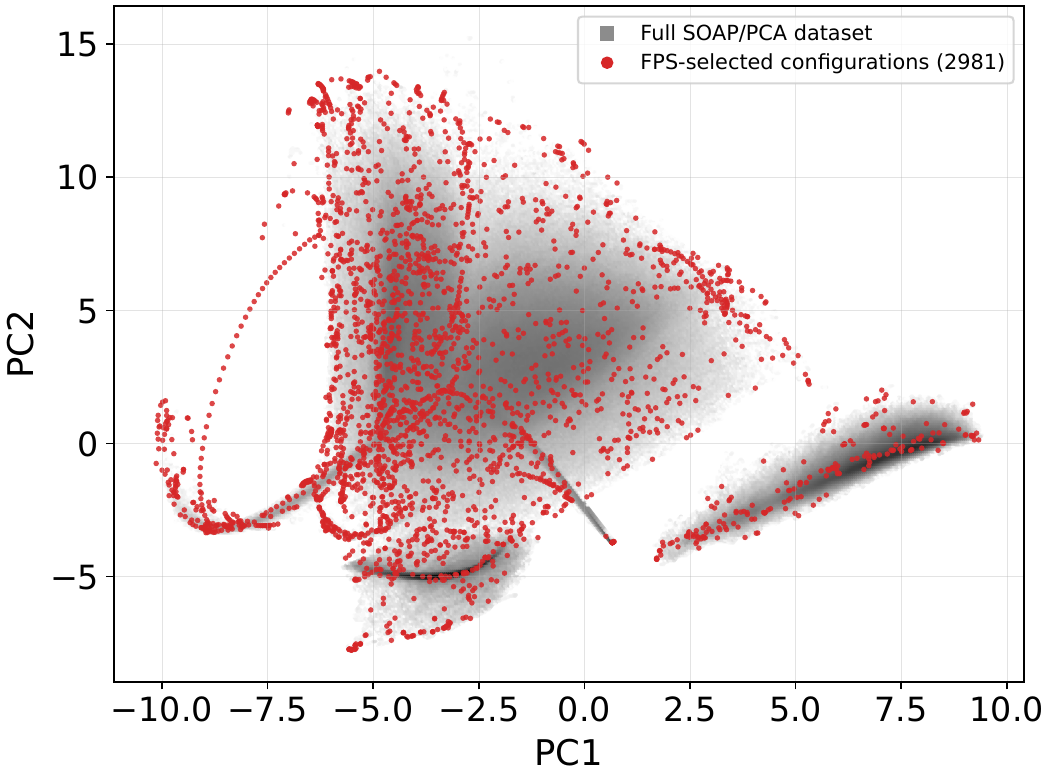}
\par\medskip\refstepcounter{figure}\label{fig:mlip-fps}
{\small\rmfamily\justifying\noindent\textbf{Figure~\thefigure:} PCA projection of the SOAP descriptor space for the full AIMD dataset, with FPS-selected environments highlighted as red dots.\par}

\end{minipage}
\par\medskip

Once the four Deep Potential models were trained using the configurations selected through the PCA/FPS protocol, their performance was evaluated against the same independent DFT test set of 2000 configurations. Total-energy RMSEs range from 0.0456 to \mbox{$0.0505\,\text{eV}$}, while force-magnitude RMSEs range from 0.0303 to \mbox{$0.0335\,\text{eV}\,\text{\AA}^{-1}$}. These low prediction errors are particularly encouraging considering that the test set spans an energy range of approximately 6 eV and force magnitudes up to 8 eV \AA$^{-1}$. In both cases, the RMSEs represent less than 1\% of the corresponding range, highlighting the accuracy of the models across the sampled configurations. The four models show comparable accuracy on this test set; however, model 2 yields the lowest errors for both quantities and was therefore selected for subsequent calculations. Fig.~\ref{fig:mlip-parity} presents correlation plots between MLIP model 2 predictions and DFT reference test data.

\par\medskip
\noindent\begin{minipage}{\columnwidth}
\centering
\includegraphics[width=\linewidth]{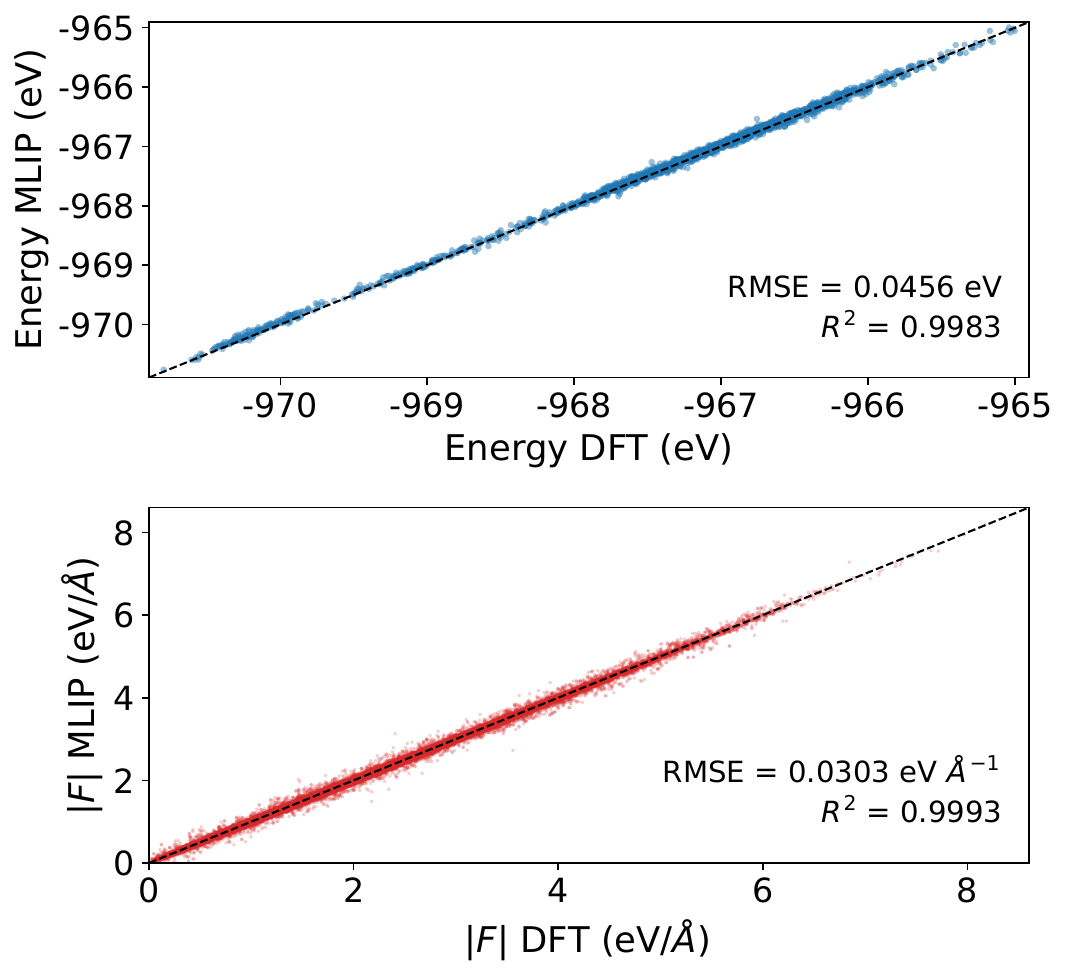}
\par\medskip\refstepcounter{figure}\label{fig:mlip-parity}
{\small\rmfamily\justifying\noindent\textbf{Figure~\thefigure:} Parity plots for the selected MLIP:
total energies (top) and all force vectors magnitudes (bottom), compared with DFT.\par}

\end{minipage}
\par\medskip

Following the selection of model 2, MLIP-driven molecular dynamics trajectories were performed within the QBC framework, as described in Section~\ref{sec:mlip-methods}. The QBC analysis showed that approximately 97\% of the evaluated configurations exhibited force prediction disagreements below \mbox{$0.05\,\text{eV}\,\text{\AA}^{-1}$}, exceeding the 95\% threshold established in the methodology. This high degree of consistency among the committee models indicated that a further active-learning iteration was not necessary for the exploratory applications considered in this work.

Overall, although only one reactive event was observed during the exploratory AIMD sampling, limiting the representation of configurations associated with NO oxidation, the developed MLIP provides an accurate description of the sampled non-reactive regions of the configurational space. This makes it a valuable tool for systematically exploring the NO--surface interaction landscape and identifying configurations of interest for further first-principles investigations.

\subsection{Applications of the selected MLIP}
\label{sec:mlip-applications}

The selected MLIP was subsequently applied to investigate the NO--O$_{16}$@HOPG interaction landscape. In particular, adsorption sampling and spatial and angular interaction energy maps provide insight into favorable molecular configurations and help identify the key factors governing NO--surface interactions, including those potentially relevant to NO oxidation.

The computational efficiency of this approach was also assessed using the 20 MLIP-driven trajectories propagated with LAMMPS \cite{Thompson2022LAMMPS} during the QBC analysis, as described in Section~\ref{sec:mlip-methods}. For a collision trajectory of up to 4 ps, the computational cost was reduced from approximately 3500 CPU-hours for AIMD to only 2 CPU-hours for MLIP-driven dynamics. This approximately 1600-fold reduction in computational cost enables substantially more extensive configurational sampling, facilitating the systematic exploration of NO--surface interactions.

\subsubsection{Exploration of the NO adsorption landscape}

\begin{figure*}[pos=t!]
\centering
\includegraphics[width=0.90\textwidth]{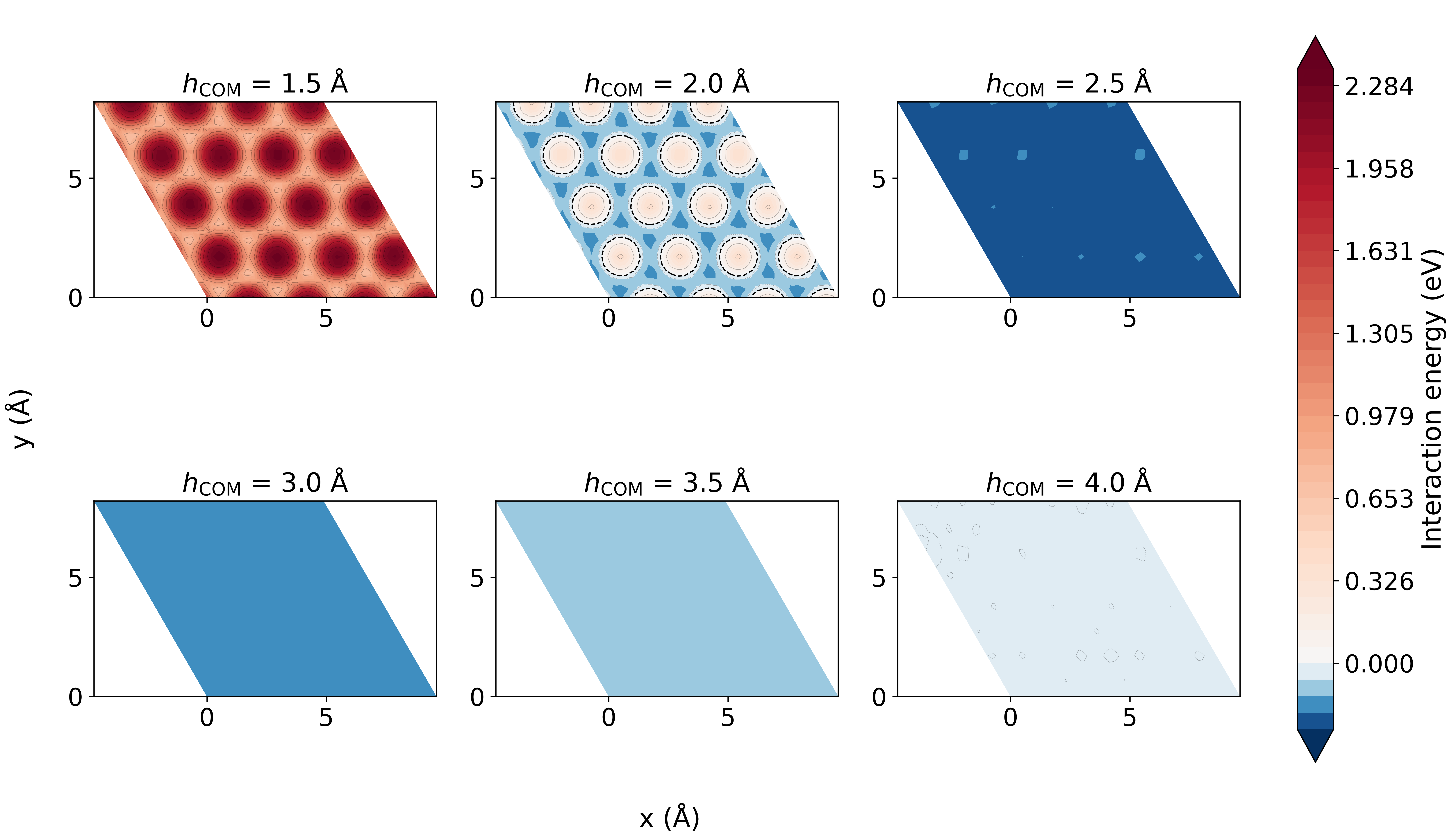}
\par\medskip\refstepcounter{figure}\label{fig:mlip-lateral}
{\small\rmfamily\justifying\noindent\textbf{Figure~\thefigure:} Spatial interaction-energy maps for vertical N-down NO above O$_{16}$@HOPG, calculated using the selected MLIP. Each panel shows the interaction energy as a function of the lateral position ($x$, $y$) at a fixed NO center-of-mass height ($h_{\mathrm{COM}}$) relative to the mean height of the surface oxygen atoms. Blue and red regions indicate attractive and repulsive interactions, respectively.\par}
\end{figure*}

Exploring the adsorption landscape requires relaxing many different initial molecular placements. Performing hundreds of such optimizations directly with DFT would be computationally demanding; the selected MLIP makes this broader search feasible at substantially lower cost.

Here, 300 configurations with random NO positions, heights, and orientations were relaxed, extending the exploration beyond the 15 configurations considered in the initial DFT study. All optimizations converged to residual forces below \mbox{$0.05\,\text{eV}\,\text{\AA}^{-1}$}.

The lowest adsorption energy obtained from the MLIP geometry optimizations was \mbox{$-0.48\,\text{eV}$}, in excellent agreement with the non-spin-polarized DFT minimum of \mbox{$-0.49\,\text{eV}$}. Both energies were calculated using the same near--far reference convention, with method-specific reference geometries. The non-spin-polarized DFT result provides a consistent comparison, as the MLIP was trained on data obtained using the same spin treatment.

These results demonstrate the ability of the selected MLIP to identify energetically favorable NO adsorption geometries. This provides an efficient approach to systematically exploring the adsorption configurational space and identifying the most favorable molecular arrangements at the oxygen-covered surface.

\subsubsection{Spatial and angular interaction maps}

A further application of the selected MLIP was the construction of interaction maps over a grid of 1000 points above the O$_{16}$@HOPG surface. This sampling provides a spatial description of the interaction, allowing attractive and repulsive regions to be identified as NO approaches the surface.

Fig.~\ref{fig:mlip-lateral} shows the maps for vertical N-down NO above the unthermalized surface. Blue indicates attractive interactions and red indicates repulsive interactions, with the same colour scale used at all molecule--surface distances. At distances of \mbox{$2.5\text{--}4.0\,\text{\AA}$}, the interaction is attractive and exhibits only small variations across the surface, allowing NO to approach the surface from nearly any position above the oxygenated layer.

At \mbox{$2.0\,\text{\AA}$}, attractive and repulsive regions coexist, with the latter becoming increasingly prominent and restricting the accessible configurational space to smaller regions of the surface. At \mbox{$1.5\,\text{\AA}$}, all sampled positions become repulsive, indicating that the oxygenated layer creates a substantial energetic barrier to the close approach of NO in the N-down orientation. This behavior suggests that the high density of surface oxygen atoms may significantly constrain the molecular approach geometries potentially favorable for NO oxidation. In contrast, in the previously studied O@HOPG system, the presence of a single isolated epoxy oxygen atom leaves a larger fraction of the graphite surface free from oxygen functionalization \cite{Alou2024}, potentially allowing greater configurational accessibility during NO collisions. Although a direct comparison of the interaction energy landscapes would be necessary to quantify this effect, the present results suggest that increasing oxygen coverage may restrict access to potentially reactive configurations, despite providing more oxygen atoms available for reaction.

This restricted accessibility may also contribute to the scarcity of reactive events observed in the exploratory AIMD trajectories. Interestingly, the only reactive trajectory followed an indirect mechanism, in which NO initially rebounded from the surface before approaching again and forming NO$_2$. Such an indirect encounter may allow the molecule to explore different orientations and redistribute its energy before reaching a reactive configuration. This observation suggests that molecular reorientation could play an important role in overcoming the geometrical and energetic constraints imposed by the oxygenated surface, motivating a more detailed examination of the angular dependence of the interaction.

\begin{figure*}[pos=t!]
\centering
\includegraphics[width=0.92\textwidth]{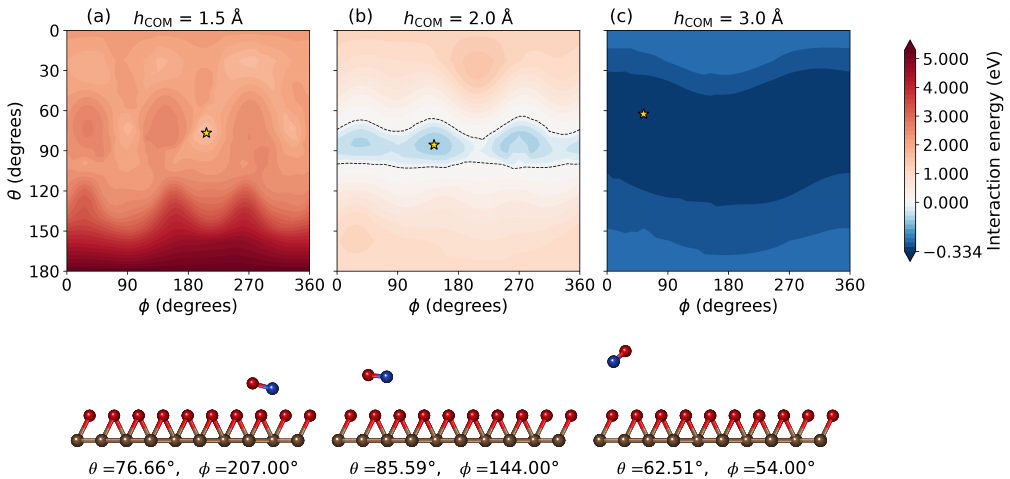}
\par\medskip\refstepcounter{figure}\label{fig:mlip-angular}
{\small\rmfamily\justifying\noindent\textbf{Figure~\thefigure:} Angular interaction-energy maps for NO above O$_{16}$@HOPG, calculated using the selected MLIP at fixed center-of-mass heights ($h_{\mathrm{COM}}$) of (a) 1.5, (b) 2.0, and (c) 3.0~\AA, measured relative to the mean height of the surface oxygen atoms. The upper panels show the interaction energy as a function of the molecular orientation angles ($\theta$, $\phi$), with blue and red indicating attractive and repulsive interactions, respectively. Yellow stars mark the lowest-energy sampled orientations, illustrated by the corresponding molecular configurations in the lower panels.\par}
\end{figure*}

The analysis therefore examined how molecular reorientation affects the approach of NO to the oxygenated surface. The angular maps reveal a progressive restriction of the energetically accessible configurational space as NO approaches the surface. The lowest-energy sampled orientations, marked by stars in the maps, correspond to the configurations shown in the lower row of Fig.~\ref{fig:mlip-angular}. At a center-of-mass height of \mbox{$3.0\,\text{\AA}$}, all sampled orientations are attractive, with a minimum of approximately \mbox{$-0.33\,\text{eV}$}. At this distance, NO can explore a broad range of molecular orientations without encountering significant energetic restrictions. However, as the molecule approaches the oxygenated layer, repulsive interactions become increasingly important, considerably reducing the range of favorable orientations. At \mbox{$2.0\,\text{\AA}$}, only a narrow band of nearly horizontal orientations remains attractive, with a minimum of approximately \mbox{$-0.12\,\text{eV}$} at \mbox{$\theta=85.59^\circ$}. This indicates that the approach of NO becomes strongly constrained by its molecular orientation, with only a limited region of the sampled configurational space remaining energetically favorable (Fig.~\ref{fig:mlip-angular}).

At even shorter distances, these restrictions become more pronounced. At a center-of-mass height of \mbox{$1.5\,\text{\AA}$}, all sampled orientations are repulsive, including the lowest-energy configuration, which exhibits an interaction energy of approximately \mbox{$1.58\,\text{eV}$} at \mbox{$\theta=76.66^\circ$} and \mbox{$\phi=207^\circ$}. Although this configuration represents a minimum within the angular map, it does not correspond to an energetically favorable approach pathway. Under low-energy collision conditions, NO would be expected to undergo scattering or reorientation before reaching such strongly repulsive configurations. The progressive disappearance of attractive regions therefore suggests that the densely oxygenated surface imposes substantial geometric and energetic constraints on the approach of NO, potentially limiting access to configurations favorable for oxygen abstraction.

Beyond the energetic accessibility of these configurations, the relative orientation of NO and the surface oxygen atoms also plays an important role in the possible formation of NO$_2$. In particular, the O$_{\mathrm{NO}}$--N--O$_{\mathrm{surf}}$ angle provides a useful geometrical descriptor of the approach toward the bent NO$_2$ product, whose equilibrium O--N--O angle is approximately 134$^\circ$. Interestingly, at a center-of-mass height of \mbox{$2.0\,\text{\AA}$}, where attractive configurations remain accessible, molecular reorientation brings NO toward geometries approaching this characteristic angle. This suggests that the energetically favorable orientations may also be compatible with NO$_2$ formation. Previous AIMD simulations on O@HOPG showed that NO$_2$ formation through oxygen abstraction is highly exothermic, releasing approximately 1.6 eV \cite{Alou2024}. Although this energetic result cannot be directly extrapolated to the highly oxygenated surface, it highlights the thermodynamic driving force associated with NO$_2$ formation in the reference system. These results suggest that, despite the larger number of potentially reactive oxygen atoms, the strong short-range repulsion imposed by the highly oxygenated surface may hinder NO oxidation by restricting access to favorable reaction geometries.

Overall, these maps demonstrate the ability of the MLIP approach to systematically explore the spatial and orientational dependence of the NO--surface interaction at a computational cost that would be prohibitive using DFT alone. Beyond providing a more detailed understanding of the interaction landscape, they offer valuable guidance for selecting additional first-principles configurations to improve the description of potentially reactive regions and support future large-scale molecular dynamics simulations.

\section{Conclusions}

A limited set of exploratory AIMD simulations was combined with machine-learning interatomic potentials to investigate NO interactions with highly oxygen-covered graphite. The first-principles configurations provided the basis for representative dataset selection, MLIP training, and validation against independent DFT data.

The exploratory AIMD trajectories revealed that short NO--surface oxygen contacts can occur without oxygen abstraction. The observed NO$_2$-forming trajectory involved an initial rebound followed by a reactive encounter, suggesting that molecular approach geometry, together with proximity to a surface oxygen atom, influences product formation.

The four trained MLIPs showed comparable agreement with the DFT test data, and the selected model enabled relaxation of 300 initial adsorption configurations and construction of interaction maps. The spatial maps reveal a progressive restriction of attractive regions as vertical N-down NO approaches the surface. The angular maps show that near-horizontal NO orientations remain attractive at \mbox{$2.0\,\text{\AA}$}, while molecular rotation reduces repulsion at \mbox{$1.5\,\text{\AA}$}. Together, these maps identify positions and orientations for targeted first-principles sampling. The additional molecular dynamics calculations required minutes rather than days per trajectory, making more extensive configurational sampling computationally feasible.

Further development will combine map-guided sampling of potentially reactive configurations with MLIP refinement and validation in the oxygen-abstraction region. Fine-tuning pretrained foundation models for this gas--surface system will also be explored. These steps will support large-scale reactive molecular dynamics simulations aimed at resolving the competition between scattering, trapping, and NO oxidation on oxygen-covered graphite.

\section{Acknowledgments}

The French State under the France-2030 programme and the Initiative of Excellence of the University of Lille are acknowledged for the funding and support granted to the R-CDP-24-003-AREA project.

The authors acknowledge the computational resources provided by the French national supercomputing facilities (Grant No. DARI A0190801859) and the Centre de Ressources Informatiques (CRI) of the Université de Lille.

\bibliographystyle{elsarticle-num}
\bibliography{cas-refs}

\end{document}